\documentstyle[prc,aps,epsfig,multicol]{revtex}
\tightenlines

\begin{document}
\title{Charm production from photo-nucleon reaction in a hadronic model}
\bigskip
\author{Wei Liu$^1$, Su Houng Lee$^{1,2}$, and Che Ming Ko$^1$}
\address{$^1$Cyclotron Institute and Physics Department, Texas A\&M University,
College Station, Texas 77843-3366, USA \\
$^2$ Department of Physics and Institute of Physics and Applied
Physics, Yonsei University, Seoul 120-749, Korea}
\maketitle

\begin{abstract}
We study the total cross section for photo production of charmed
hadrons near threshold using a hadronic Lagrangian. Both two-body
final states involving $\Lambda_c$ and a charmed meson as well as
three-body final states involving nucleon and a charm-anticharm
meson pair are included. With appropriate cut-off parameters in
the form factors at interaction vertices, the model gives a total
cross section that is consistent with the measured data at
center-of-mass energy of 6 GeV. The result is compared with the
prediction from the leading-order perturbative QCD.

\end{abstract}

\begin{multicols}{2}

\section{introduction}

Reliable estimates of the production and scattering cross sections of
open and/or hidden charmed hadrons in hadronic matter are important
for understanding many phenomena in relativistic heavy ion collisions
and hadron-nucleus reactions. In particular, the dissociation cross
sections of $J/\psi$ by light meson or nucleon are directly related
to the role of hadronic suppression of $J/\psi$ production in heavy
ion collisions \cite{Ko1,others}, and have thus been a subject under
active investigations. In the heavy charm quark mass limit, a formula
based on the leading-order (LO) perturbative QCD was derived in Refs.
\cite{Peskin1,Peskin2} for the absorption cross sections of
$J/\psi$ by hadrons. Higher-order corrections due to the target mass
and relativistic effects have also been estimated
\cite{Kharzeev1,Arleo02,Oh02}. However, the perturbative QCD approach
is useful only at very high energy when nontrivial higher twist and
higher $\alpha_s$ corrections are small. At low energies, these
corrections become large, so nonperturbative approaches are needed.
Various phenomenological models have thus been introduced for studying
the $J/\psi$ absorption cross section by hadrons at low energies.
These include the hadronic model based on effective Lagrangians
\cite{hadronic,liu2,Marina2}, the QCD sum rules \cite{Marina1}, and
the quark-exchange model \cite{Wong}.  Results from these phenomenological
models all give much larger cross sections for $J/\psi$ absorption
by light mesons than that given by the perturbative QCD formula.
To test the prediction from the hadronic model, the same effective
Lagrangian has been used to evaluate the absorption cross section
of $J/\psi$ by nucleon, and it is found to be consistent with that
extracted from photonuclear production of $J/\psi$ \cite{liu2}.

In the present paper, we generalize the effective hadronic
Lagrangian, which has also been used to study charmed meson scattering
by hadrons \cite{lin,di} and charmed meson production from meson-nucleon
scattering \cite{liu1}, to include the photon and to study charmed hadron
production from photo-nucleon reaction near threshold. Both two-body
($\Lambda_c D$) and three-body ($ND\bar D$) final states are included.
We find that using reasonable values for the cutoff parameters at the
interaction vertices, the resulting charmed hadron production cross
section near threshold energy is consistent with the measured one
at center-of-mass energy of 6 GeV \cite{slac}, although at high energies
its value is much smaller than that measured experimentally or given by
the LO perturbative QCD. We further find that the relative contribution
of two-body to three-body final states is also consistent with the
experimental data. Our results thus provide another independent test
and confirmation of the validity of hadronic model for determining the
cross sections involving heavy flavors at low energies. The effective
hadronic Lagrangian used in the present study will also be useful for
evaluating other low energy cross sections involving heavy flavored
hadrons, which can be studied at both the Japanese Hadron facility
\cite{JPARC} and at the GSI future accelerator \cite{GSI-future},
where open and hidden charmed hadrons are copiously produced
in proton- and antiproton-nucleus reactions near threshold.

This paper is organized as follows. In Section \ref{three}, we first
calculate the cross sections for photoproduction of charmed hadrons
from nucleon with three particles in the final state, which is dominated
by pion and rho meson exchange. Photoproduction of charmed hadrons
from nucleon with two particle in the final state, which includes
the charmed meson exchange, is studied in Section \ref{two}. In Section
\ref{total}, we show the total cross section and compare it to available
experimental data and results from the LO QCD. Finally, summary and
discussions are given in Section \ref{conclusion}.

\section{Photoproduction of charmed hadrons from nucleon with
three-particle final states}
\label{three}

\begin{figure}[ht]
\centerline{\epsfig{file=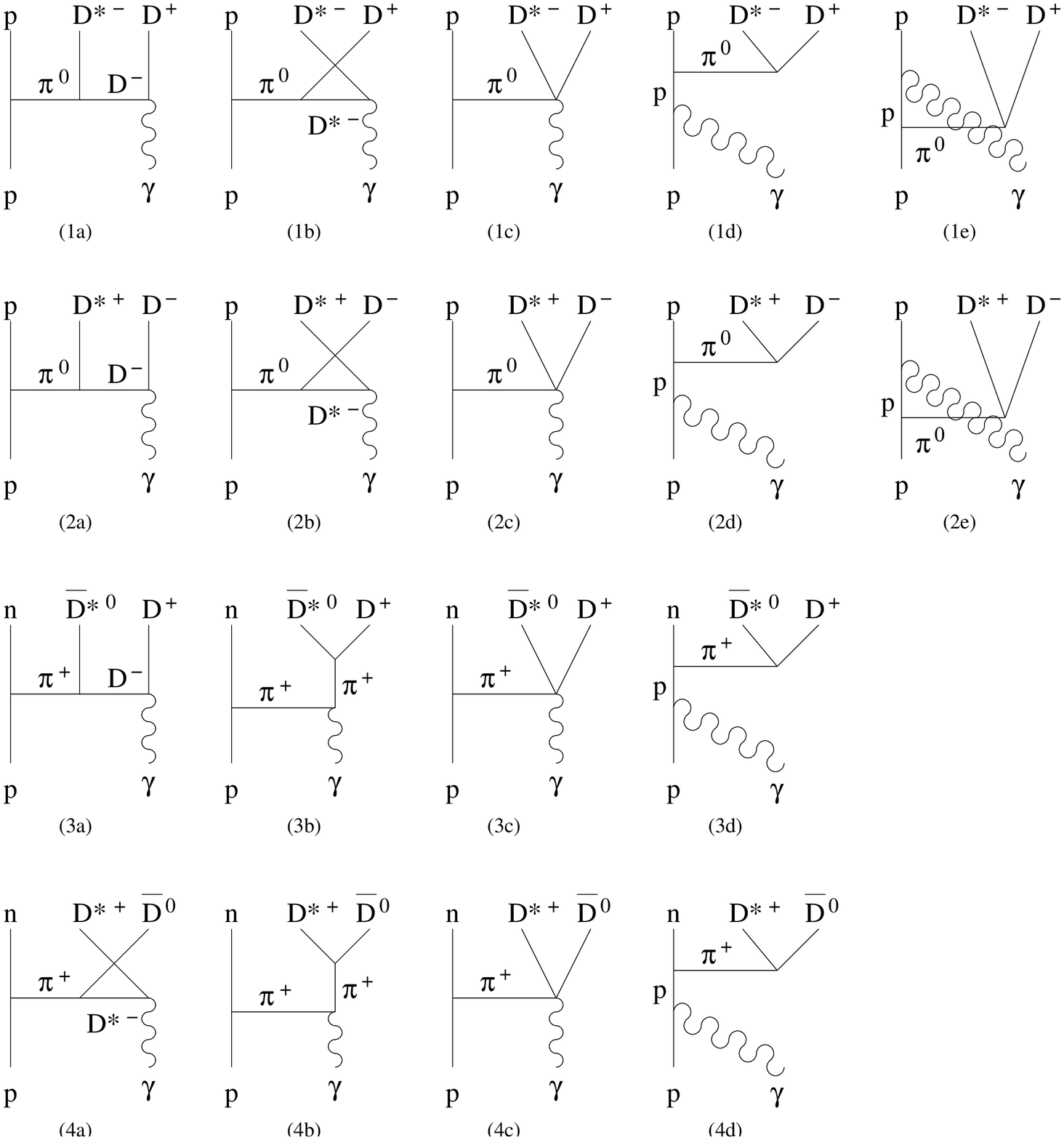,width=3in,height=4.0in,angle=0}}
\vspace{0.5cm}
\caption{Photoproduction of charmed hadrons from nucleon involving
pion exchange.}
\label{figure1}
\end{figure}

\begin{figure}[ht]
\centerline{\epsfig{file=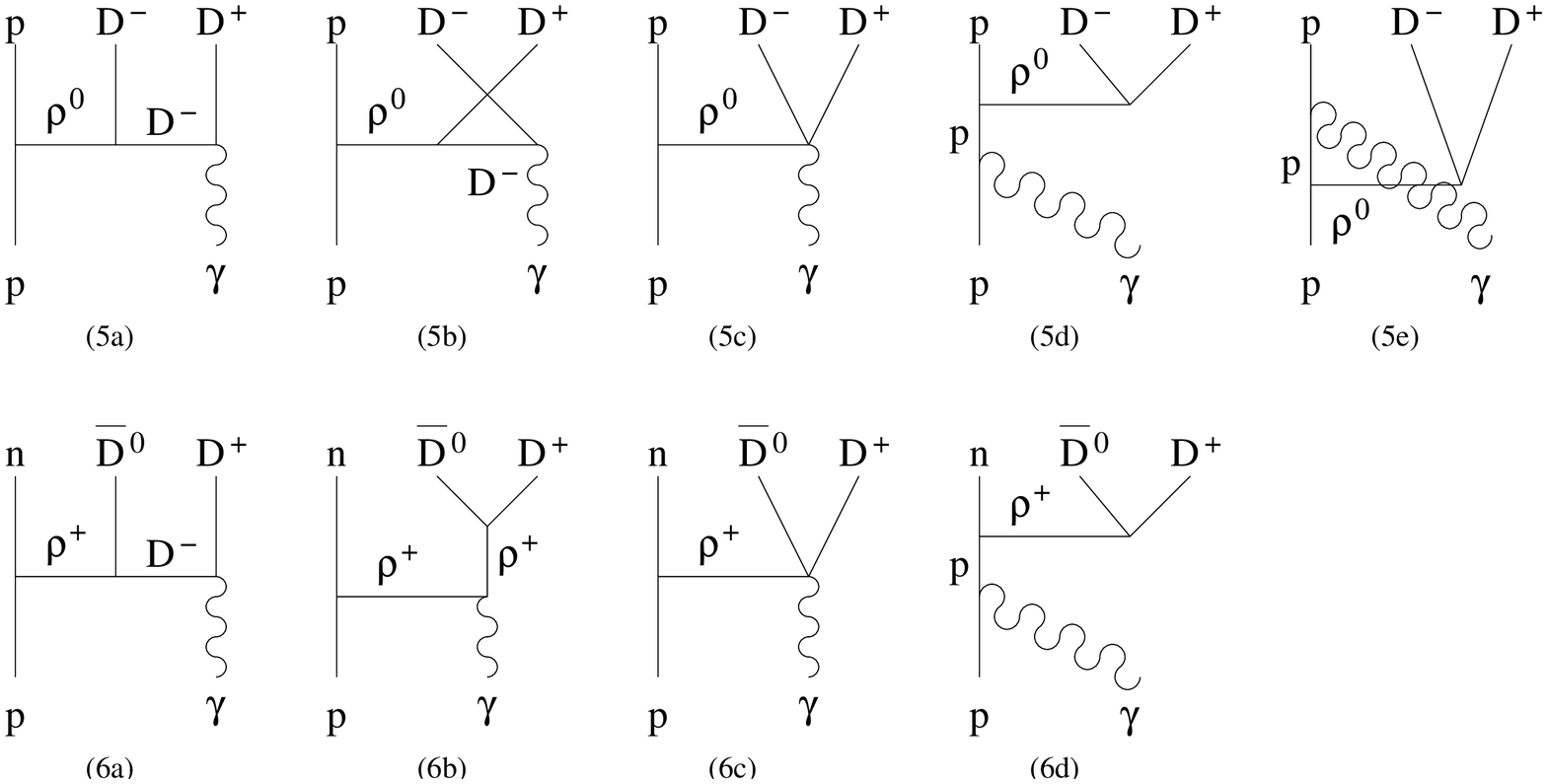,width=3in,height=2.0in,angle=0}}
\vspace{0.5cm}
\caption{Photoproduction of charmed hadrons ($D\bar D)$ from proton
involving rho meson exchange.}
\label{figure2}
\end{figure}

\begin{figure}[ht]
\centerline{\epsfig{file=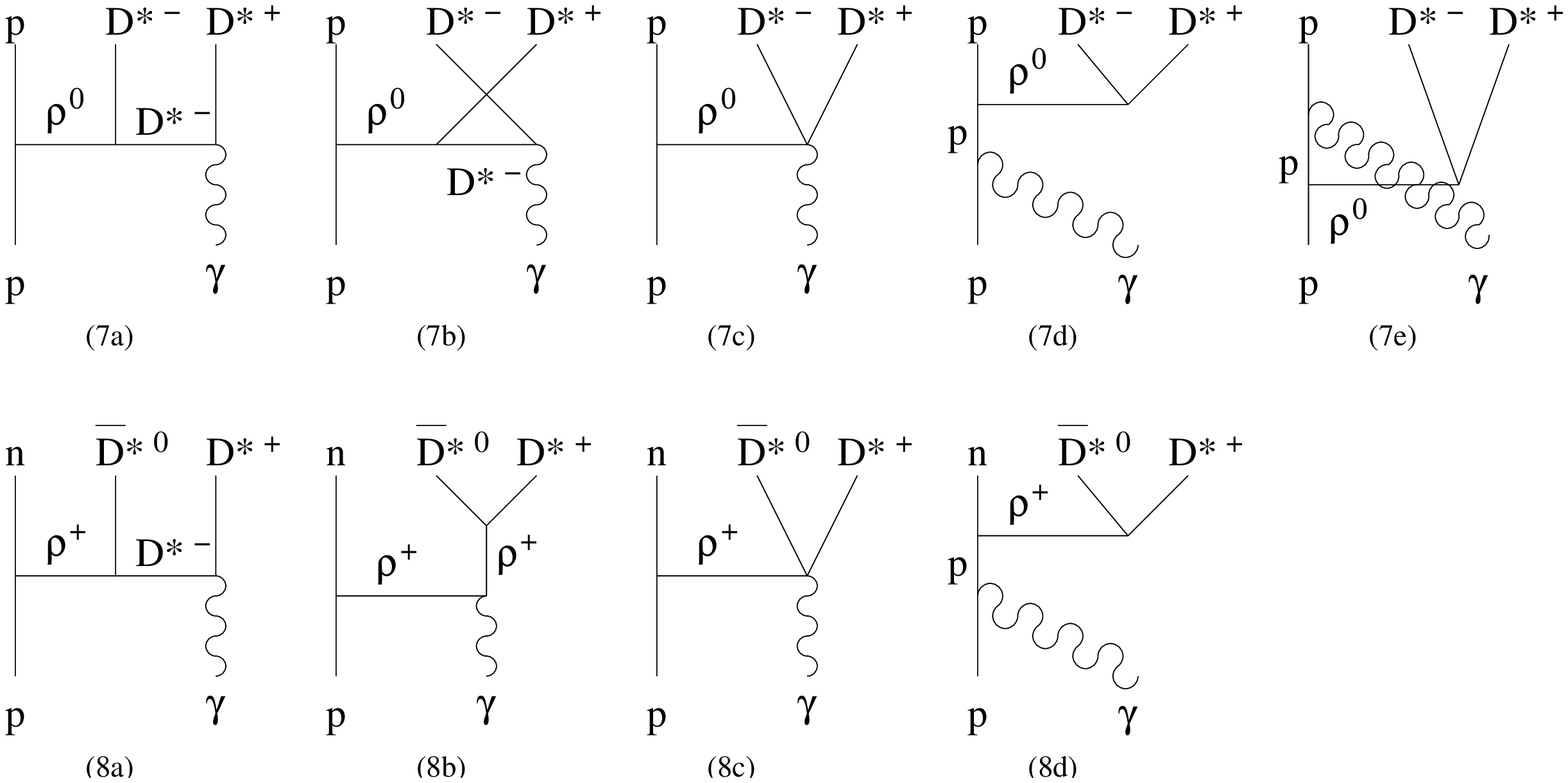,width=3in,height=2.0in,angle=0}}
\vspace{0.5cm}
\caption{Photoproduction of charmed hadrons ($D^*\bar D^*$) from proton
involving rho meson exchange.}
\label{figure3}
\end{figure}

In photoproduction of charmed hadrons from nucleon with three particle
in the final states, the three possible reactions are
$\gamma N\rightarrow \bar{D}DN$,
$\gamma N\rightarrow\bar{D}D^*N(\bar{D}^*DN)$ ,and
$\gamma N\rightarrow\bar{D}^*D^*N$. The lowest-order diagrams for the
process $\gamma N\rightarrow \bar{D}DN$ are shown in Fig. \ref{figure1}
and involves the exchange of pion in the intermediate state; those for
the processes $\gamma N\rightarrow\bar{D}D^*N(\bar{D}^*DN)$ and
$\gamma N\rightarrow\bar{D}^*D^*N$ involve the exchange of rho meson,
and the lowest-order diagrams for the two processes are shown in
\ref{figure2} and \ref{figure3}.

To evaluate the cross sections for these processes, we use the same
Lagrangian introduced in Refs. \cite{liu2,di,liu1} for studying charmed
meson scattering by hadrons. This Lagrangian is based on the gauged SU(4)
flavor symmetry but with empirical masses. The coupling constants are
taken, if possible, from empirical information. Otherwise, the
SU(4) relations are used to relate the unknown coupling constants
to the known ones.  Photon is then introduced in the Lagrangian via
gauging its U$_{\rm em}(1)$ part as in ref.\cite{Song}.

The interaction Lagrangian densities that are relevant to the processes
shown in Figs. \ref{figure1}, \ref{figure2}, and \ref{figure3} are given
as follows:
\begin{eqnarray}
{\cal L}_{\pi NN} & = & -ig_{\pi NN}\bar{N}\gamma_{5}
\vec{\tau}N\cdot\vec{\pi},\nonumber\\
{\cal L}_{\rho NN} & = & g_{\rho NN}\bar{N}\left(\gamma^{\mu}\vec{\tau}\cdot
\vec{\rho}_{\mu}+\frac{\kappa_{\rho}}{2m_{N}}\sigma_{\mu\nu}\vec{\tau}
\cdot\partial_{\mu}\vec{\rho}_{\nu}\right)N, \nonumber\\
{\cal L}_{\pi DD^{*}} & = & ig_{\pi DD^{*}}D^{*\mu}\vec{\tau}\cdot(\bar{D}
\partial_{\mu}\vec{\pi}-\partial_{\mu}\bar{D}\vec{\pi})+{\rm H.c.},
\nonumber\\
{\cal L}_{\rho DD} & = & ig_{\rho DD}(D\vec{\tau}\partial_{\mu}\bar{D}-
\partial_{\mu}D\vec{\tau}\bar{D})\cdot\vec{\rho}^{\mu},\nonumber\\
{\cal L}_{\rho D^{*}D^{*}} & = & ig_{\rho D^{*}D^{*}}[(\partial_{\mu}D^{*\nu}
\vec{\tau}\bar{D}^{*}_{\nu}-D^{*\nu}\vec{\tau}\partial_{\mu}
\bar{D}^{*}_{\nu})\cdot\vec{\rho}^{\mu} \nonumber\\
&+&(D^{*\nu}\vec{\tau}\cdot\partial_{\mu}\vec{\rho}^{\nu}
-\partial_{\mu}D^{*\nu}\vec{\tau}\cdot\vec{\rho}_{\nu})\bar{D}^{*\mu}
\nonumber\\
&+&D^{*\mu}(\vec{\tau}\cdot\vec{\rho}^{\nu}\partial_{\mu}
\bar{D}^{*}_{\nu}-\vec{\tau}\cdot\partial_{\mu}\vec{\rho}^{\nu}
\bar{D}^{*}_{\nu})],\nonumber\\
{\cal L}_{\gamma DD} & = & ie A^{\mu}[D Q \partial_\mu \bar D
-(\partial_{\mu}D)Q\bar D],\nonumber\\
{\cal L}_{\gamma D^{*}D^{*}} & = & ie[A^{\mu}
(\partial_{\mu}D^{*\nu}Q\bar{D}^{*}_{\nu}-D^{*\nu}Q\partial_{\mu}
\bar{D}^{*}_{\nu}) \nonumber\\
&+&(\partial_{\mu}A^\nu D^{*}_{\nu}-A^{\nu}
\partial_{\mu}D^{*}_{\nu})Q\bar{D}^{*\mu}\nonumber\\
&+&D^{*\mu}Q(A^{\nu}
\partial_{\mu}D^{*}_{\nu}-\partial_{\mu}A^{\nu}\bar{D}^{*}_{\nu}],
\nonumber\\
{\cal L}_{\pi \gamma DD^{*}} & = & -e g_{\pi DD^{*}}A^{\mu}(D^{*}_{\mu}
(2\vec{\tau}Q-Q\vec{\tau}) \bar{D}\nonumber\\
&+&D(2Q\vec{\tau}-\vec{\tau} Q)\bar{D}^{*}_{\mu})\cdot\vec{\pi},\nonumber\\
{\cal L}_{\rho \gamma DD} & = & e g_{\rho DD}A^{\mu}D(\vec{\tau}Q
+ Q \vec{\tau})\bar{D}\cdot\vec{\rho}_{\mu}, \nonumber\\
{\cal L}_{\rho \gamma D^{*}D^{*}} & = & e g_{\rho D^{*}D^{*}}
(A^{\nu}D^{*}_{\nu}(2 \vec{\tau}Q-Q\vec{\tau})\bar{D}^{*}_{\mu}\nonumber\\
&+&A^{\nu}D^{*}_{\mu}( 2\vec{\tau}Q-Q\vec{\tau}) \bar{D}^{*}_{\nu} \nonumber\\
&-&A_{\mu}D^{*\nu}(2 \vec{\tau} Q -Q\vec{\tau} )
\bar{D}^{*}_{\nu})\cdot\vec{\rho}^{\mu}.
\end{eqnarray}
In the above, ${\vec\tau}$ are Pauli spin matrices, and
$\vec{\pi}$ and $\vec{\rho}$ denote the pion and rho meson isospin
triplet, respectively, while $D=(D^0,D^+)$ and
$D^*=({D^*}^0,{D^*}^+)$ denote the pseudoscalar and vector charm
meson doublets, respectively.  The operator $Q$ is the diagonal charge
operator with diagonal elements equal to 0 and -1.

For coupling constants in the above interaction Lagrangians, we use the
following empirical values: $g_{\pi NN}=13.5$ \cite{pnn}, $g_{\rho NN}=3.25$,
$\kappa_\rho=6.1$ \cite{rnn}, $g_{\pi DD^*}=5.56$ \cite{liu1,pdd}, and
$g_{\rho DD}=2.52$ \cite{di}.

The diagrams in Figs. \ref{figure1}, \ref{figure2}, and \ref{figure3}
can be separated to two types; one in which the photon is coupled to mesons
such as the first three diagrams (denoted by (ia) to (ic) with i=1 to 8),
and the other in which the photon is coupled directly to the incoming or
outgoing charged nucleon.  As shown later, contributions from the latter
type are much smaller than those from the first type of diagrams and
are thus neglected in the following calculations. We note that diagrams
of the first type are similar to those for $J/\psi$ absorption on nucleon,
which can be interpreted as absorption by the virtual pion and rho meson
cloud of a nucleon. Here, they can be considered as charmed hadron
production from the nucleon's virtual meson cloud.

The amplitudes for the four processes in Fig. \ref{figure1} are given by
\begin{eqnarray}
{\cal M}_i & = & -iag_{\pi NN}\bar{N}(p_{3})\gamma_{5}N(p_{1})
\frac{1}{t-m^{2}_{\pi}}\nonumber\\
&\times&(M_{ia}+M_{ib}+M_{ic}),
\end{eqnarray}
with $i=1$ to 4, while the amplitudes for the four processes in
Fig. \ref{figure2}, \ref{figure3} can be written as
\begin{eqnarray}
{\cal M}_{j} & = &ag_{\rho NN}\bar{N}(p_{3})
\left[\gamma^{\mu}
+i\frac{\kappa_{\rho}}{2m_{N}}\sigma^{\alpha\mu}
(p_{1}-p_{3})_{\alpha}\right]\nonumber\\
&\times& N(p_{1})\left[-g_{\mu\nu}+\frac{(p_1-p_3)_{\mu}
(p_1-p_3)_{\nu}}{m^{2}_{\rho}}\right]\nonumber\\
&\times&\frac{1}{t-m^{2}_{\rho}}
(M^{\nu}_{ja}+M^{\nu}_{jb}+M^{\nu}_{jc}),
\end{eqnarray}
with $j=5$ to 8. In the above, $p_1$ and $p_3$ are four momenta of the
initial and final nucleons, respectively. The coefficient $a$ is,
respectively, 1 and $\sqrt 2$ for neutral and charged pion coupling to
nucleon.

The three amplitudes $M_{ia}$, $M_{ib}$, and $M_{ic}$ are for the
subprocess $\pi \gamma \to D^{*}\bar{D}$ in Fig. \ref{figure1}. It
can be shown that they fulfill the chiral constraint \cite{Marina2},
i.e., their sum vanishes at the soft pion limit. Explicitly, they are
given by
\begin{eqnarray}\label{photopion}
{\cal M}_{1a}&=&{\cal M}_{2a}\nonumber\\
&=&eg_{\pi DD^*}(-2k_1 +k_3)^\mu\frac{1}{t-m^2_D}\nonumber\\
&\times&(k_1 -k_3 +k_4)^\nu\varepsilon_{3\mu}\varepsilon_{2\nu},\nonumber\\
{\cal M}_{1b}&=&{\cal M}_{2b}\nonumber\\
&=&-eg_{\pi DD^*}(-k_1-k_4)^\alpha\frac{1}{u-m^2_{D^*}}
\nonumber\\
&\times&\left[g_{\alpha\beta}-\frac{(k_1-k_4)_\alpha(k_1-k_4)_\beta}
{m^2_{D^*}}\right]\nonumber\\
&\times&[(-k_2 -k_3)^\beta g^{\mu\nu}+(-k_1+k_2 +k_4)^\nu g^{\beta\mu}
\nonumber\\
&+&(k_1+k_3-k_4)^\mu g^{\beta\nu}]\varepsilon_{3\mu}\varepsilon_{2\nu},
\nonumber\\
{\cal M}_{1c}&=&{\cal M}_{2c}\nonumber\\
&=&eg_{\pi DD^*}g^{\mu\nu}\varepsilon_{3\mu}
\varepsilon_{2\nu},\nonumber\\
{\cal M}_{3a}&=&\sqrt{2}eg_{\pi DD^*}(-2k_1 +k_3)^\mu\frac{1}{t-m^2_D}
\nonumber\\
&\times&(k_1 -k_3 +k_4)^\nu\varepsilon_{3\mu}\varepsilon_{2\nu},\nonumber\\
{\cal M}_{3b}&=&-\sqrt{2}eg_{\pi DD^*}(2k_1+k_2)^\nu\frac{1}{s-m^2_\pi}
\nonumber\\
&\times&(k_1+k_2+k_4)^\mu\varepsilon_{3\mu}\varepsilon_{2\nu},\nonumber\\
{\cal M}_{3c}&=&2\sqrt{2}eg_{\pi DD^*}g^{\mu\nu}\varepsilon_{3\mu}
\varepsilon_{2\nu},\nonumber\\
{\cal M}_{4a}&=&-\sqrt{2}eg_{\pi DD^*}(-k_1-k_4)^\alpha\frac{1}{u-m^2_{D^*}}
\nonumber\\
&\times&\left[g_{\alpha\beta}-\frac{(k_1-k_4)_\alpha
(k_1-k_4)_\beta}{m^2_{D^*}}\right]\nonumber\\
&\times&[(-k_2 -k_3)^\beta g^{\mu\nu}+(-k_1+k_2 +k_4)^\nu g^{\beta\mu}
\nonumber\\
&+&(k_1+k_3-k_4)^\mu g^{\beta\nu}]\varepsilon_{3\mu}\varepsilon_{2\nu},
\nonumber\\
{\cal M}_{4b}&=&\sqrt{2}eg_{\pi DD^*}(2k_1+k_2)^\nu\frac{1}{s-m^2_\pi}
\nonumber\\
&\times&(k_1+k_2+k_4)^\mu\varepsilon_{3\mu}\varepsilon_{2\nu},\nonumber\\
{\cal M}_{4c}&=&-\sqrt{2}eg_{\pi DD^*}g^{\mu\nu}\varepsilon_{3\mu}
\varepsilon_{2\nu},
\end{eqnarray}
where $k_i$ denotes the momentum of particle $i$ of each
subprocess, and $\varepsilon_{\mu}$ and $\varepsilon_{\nu}$ are
the polarization vector of $\gamma$ and $D^*$, respectively. We choose
the convention that particles 1 and 2 represent initial-state particles
while particles 3 and 4 represent final-state ones on the left and right
sides of the diagrams.

The amplitudes $M^{\nu}_{ja}$, $M^\nu_{jb}$, and $M^\nu_{jc}$ are
those for the subprocesses $\rho \gamma \to D\bar{D}$ and $\rho \gamma \to
D^*\bar{D}^*$ in Figs. \ref{figure2} and \ref{figure3}, and they are
are written explicitly as:
\begin{eqnarray}\label{photorho}
{\cal M}^\mu_{5a}&=&-eg_{\rho DD}(k_1 -2k_3)^\mu\frac{1}{t-m^2_D}\nonumber\\
&\times&(k_1-k_3+k_4)^\nu\varepsilon_{2\nu},\nonumber\\
{\cal M}^\mu_{5b}&=&-eg_{\rho DD}(-k_1+2k_4)^\mu\frac{1}{u-m^2_D}\nonumber\\
&\times&(-k_1-k_3+k_4)^\nu\varepsilon_{2\nu},\nonumber\\
{\cal M}^\mu_{5c}&=&2eg_{\rho DD}g^{\mu\nu}\varepsilon_{2\nu},\nonumber\\
{\cal M}^\mu_{6a}&=&\sqrt{2}eg_{\rho DD}(k_1 -2k_3)^\mu\frac{1}{t-m^2_D}
\nonumber\\
&\times&(k_1-k_3+k_4)^\nu\varepsilon_{2\nu},\nonumber\\
{\cal M}^\mu_{6b}&=&\sqrt{2}eg_{\rho DD}[(-2k_1-k_2)^\nu g^{\mu\alpha}
\nonumber\\
&+&(k_1+2k_2)^\mu g^{\alpha\nu}+(k_1-k_2)^\alpha g^{\mu\nu}]
\frac{1}{s-m^2_\rho}\nonumber\\
&\times&\left[g_{\alpha\beta}-\frac{(k_1+k_2)_\alpha(k_1+k_2)_\beta}{m^2_\rho}
\right]\nonumber\\
&\times&(k_3-k_4)^\beta\varepsilon_{2\nu},\nonumber\\
{\cal M}^\mu_{6c}&=&-\sqrt{2}eg_{\rho DD}g^{\mu\nu}\varepsilon_{2\nu},
\nonumber\\
{\cal M}^\mu_{7a}&=&eg_{\rho D^* D^*}[(-k_1-k_3)^\alpha g^{\mu\lambda}
+(2k_1-k_3)^\lambda g^{\alpha\mu}\nonumber\\
&+&(2k_3-k_1)^\mu g^{\alpha\lambda}]\frac{1}{t-m^2_{D^*}}\nonumber\\
&\times&\left[g_{\alpha\beta}-\frac{(k_1-k_3)_\alpha(k_1-k_3)_\beta}
{m^2_{D^*}}\right]\nonumber\\
&\times&[-2p^\omega_2 g^{\beta\nu}+(k_2+k_4)^\beta g^{\nu\omega} -2k^\nu_4
g^{\beta\omega}]\nonumber\\
&\times&\varepsilon_{2\nu}\varepsilon_{3\lambda}\varepsilon_{4\omega},
\nonumber\\
{\cal M}^\mu_{7b}&=&eg_{\rho D^* D^*}[(-2k_1+k_4)^\omega g^{\alpha\mu}
+(k_1+k_4)^\alpha g^{\mu\omega}\nonumber\\
&+&(k_1-2k_4)^\mu g^{\alpha\omega}]\frac{1}{u-m^2_{D^*}}\nonumber\\
&\times&\left[g_{\alpha\beta}-\frac{(k_1-k_4)_\alpha(k_1-k_4)_\beta}
{m^2_{D^*}}\right]\nonumber\\
&\times&[(-k_2-k_3)^\beta g^{\nu\lambda}+2k^\lambda_2 g^{\beta\nu}+2k^\nu_3
g^{\beta\lambda}]\nonumber\\
&\times&\varepsilon_{2\nu}\varepsilon_{3\lambda}\varepsilon_{4\omega},
\nonumber\\
{\cal M}^\mu_{7c}&=&eg_{\rho D^* D^*}(g^{\mu\lambda}g^{\nu\omega}
+g^{\mu\omega}g^{\nu\lambda}-2g^{\mu\nu}g^{\lambda\omega})\nonumber\\
&\times&\varepsilon_{2\nu}\varepsilon_{3\lambda}\varepsilon_{4\omega},
\nonumber\\
{\cal M}^\mu_{8a}&=&\sqrt{2}eg_{\rho D^* D^*}[(-k_1-k_3)^\alpha
g^{\mu\lambda}+(2p_1-p_3)^\lambda g^{\alpha\mu} \nonumber\\
&+&(2k_3-k_1)^\mu g^{\alpha\lambda}]\frac{1}{t-m^2_{D^*}}\nonumber\\
&\times&\left[g_{\alpha\beta}-\frac{(k_1-k_3)_\alpha(k_1-k_3)_\beta}
{m^2_{D^*}}\right]\nonumber\\
&\times&[-2p^\omega_2 g^{\beta\nu}+(k_2+k_4)^\beta g^{\nu\omega}-2k^\nu_4
g^{\beta\omega}]\nonumber\\
&\times&\varepsilon_{2\nu}\varepsilon_{3\lambda}\varepsilon_{4\omega},
\nonumber\\
{\cal M}^\mu_{8b}&=-&\sqrt{2}eg_{\rho D^*D^*}[(-2k_1-k_2)^\nu g^{\mu\alpha}
+(k_1+2k_2)^\mu g^{\alpha\nu}\nonumber\\
&+&(k_1-k_2)^\alpha g^{\mu\nu}]\frac{1}{s-m^2_\rho}\nonumber\\
&\times&\left[g_{\alpha\beta}-\frac{(k_1+k_2)_\alpha(k_1+k_2)_\beta}{m^2_\rho}
\right]\nonumber\\
&\times&[-2k^\lambda_4 g^{\beta\omega}+2k^\omega_3 g^{\beta\lambda}
+(k_4 -k_3)^\beta g^{\lambda\omega}]\nonumber\\
&\times&\varepsilon_{2\nu}\varepsilon_{3\lambda}\varepsilon_{4\omega},
\nonumber\\
{\cal M}^\mu_{8c}&=&-\sqrt{2}eg_{\rho D^* D^*}(g^{\mu\lambda}g^{\nu\omega}
-2g^{\mu\omega}g^{\nu\lambda}+g^{\mu\nu}g^{\lambda\omega})\nonumber\\
&\times&\varepsilon_{2\nu}\varepsilon_{3\lambda}\varepsilon_{4\omega}.
\end{eqnarray}

The cross sections for these reactions with three particles in the
final state can be expressed in terms of the off-shell cross
sections of the subprocesses involving two particles in the final
state.  Following the method of Ref. \cite{yao} for the reaction
$NN\to N\Lambda K$, the spin-averaged differential cross section
for the four reactions in Fig. \ref{figure1} can be written as
\begin{eqnarray}
\frac{d\sigma_{\gamma N\to ND^*\bar D}}{dtds_1} & = &
\frac{ag^{2}_{\pi NN}}
{32\pi^{2}sp^{2}_{c}}k\sqrt{s_{1}}(-t)\frac{1}
{(t-m^{2}_{\pi})^{2}}\nonumber \\
&\times&\sigma_{\pi\gamma\to D^*\bar D}(s_{1},t)|F(t)|^2,  \label{dsigma1}
\end{eqnarray}
while those for the two reactions in Fig. \ref{figure2} are
\begin{eqnarray}
\frac{d\sigma_{\gamma N\to ND\bar D}}{dtds_{1}} & = & \frac{3
ag^{2}_{\rho NN}}{64\pi^{2}sp^{2}_{c}}k\sqrt{s_{1}}
\frac{1}{(t-m^{2}_{\rho})^{2}} \left[4(1+\kappa_{\rho})^2 \right.
\nonumber \\
&\times&(-t-2m^{2}_{N})\kappa^{2}_{\rho}\frac{(4m^{2}_{N}-t)^{2}}{2m^{2}_{N}}
+4(1+\kappa_{\rho})\nonumber\\
&\times&\left .\kappa_{\rho}(4m^{2}_{N}-t)\right]
\sigma_{\rho\gamma\to D\bar D}(s_{1},t) |F(t)|^2. \label{dsigma2}
\end{eqnarray}
In the above, $p_c$ is the center-of-mass momentum of $\gamma$ and
$N$, $t$ is the squared four momentum transfer, and $s$ is the
squared center-of-mass energy.   The quantity $s_{1}$ and $k$ are,
respectively, the squared invariant mass and center-of-mass momentum
of the $\pi$ and $\gamma$ in the subprocess $\gamma N\to D^*\bar DN$ or of
the $\rho$ and $\gamma$ in the subprocesses $\gamma N\to D\bar DN$ and
$\gamma N\to D^*\bar D^*N$. Cross sections for these subprocesses
are obtained from the amplitudes in Eqs.(\ref{photopion}) and
(\ref{photorho}) using the software package FORM \cite{form} to evaluate
the summation over the polarizations of both initial and final particles.
The differential cross sections for the two reactions
$\gamma N\to D^*\bar D^*N$ in Fig. \ref{figure3} are similar to those
for $\gamma N\to D\bar DN$ with $\sigma_{\rho\gamma\to D\bar D}(s_{1},t)$
replaced by $\sigma_{\rho\gamma\to D^*\bar D^*}(s_{1},t)$.

We have introduced the form factors $F_{\pi NN}$ and $F_{\rho NN}$ at the
$\pi NN$ and $\rho NN$ vertices, respectively, to take into account the
finite size of hadrons. As in Ref. \cite{liu2}, both are taken to have
the following monopole form:
\begin{eqnarray}
F(t)=\frac{\Lambda^2-m^2}{\Lambda^2-t},
\end{eqnarray}
where $m$ is the mass of the exchanged pion or rho meson, and
$\Lambda$ is a cutoff parameter.   Following Ref.\cite{liu2} we
take $\Lambda_{\pi NN}=1.3$ GeV and $\Lambda_{\rho NN} =1.4$ GeV.
We have also introduced a universal form factor at the strong interaction
vertices in the $\pi \gamma\rightarrow D^* D$, $\rho \gamma \rightarrow DD$,
and $\rho \gamma\rightarrow D^* D^*$ two-body subprocesses. Such a
prescription guarantees the gauge invariance when all diagrams are
included in each process \cite{lee}.   The form factor we include here is
of the following dipole form:
\begin{eqnarray}
f({\bf q}^2)=\left(\frac{\Lambda^2}{\Lambda^2+{\bf q}^2}\right)^2,
\label{form}
\end{eqnarray}
with ${\bf q}$ denoting the three momentum of photon in the center-of-mass
system.  We choose the cutoff parameter $\Lambda$ to be 1.45 GeV to best
reproduce the data. This form is also our default choice of form factors
in this paper, unless stated otherwise.

\begin{figure}[ht]
\centerline{\epsfig{file=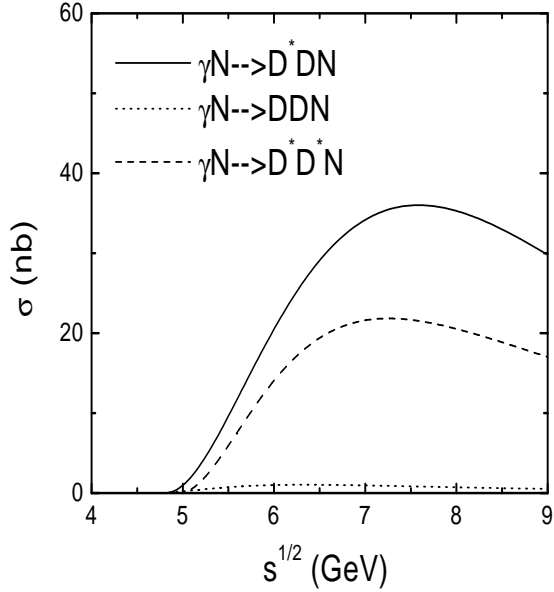,width=3.5in,height=3.5in,angle=-90}}
\caption{Cross sections for photoproduction of charmed hadrons from nucleon
with three particles in the final states.}
\label{cross1}
\end{figure}

The cross sections for charmed hadron production from neutron is
the same as that from proton if we neglect diagrams
involving photon coupling directly to proton.
The isospin-averaged cross section for the reaction
$\gamma N\rightarrow \bar{D}DN$ is thus given by the sum of the cross sections
for the four processes in Fig. \ref{figure1}, which are obtained by
integrating Eq.(\ref{dsigma1}) over $t$ and $s_1$.  Similarly, one can
obtain from Eq.(\ref{dsigma2}) the isospin averaged cross sections
for the reactions $\gamma N\rightarrow\bar{D}D^*N(\bar{D}^*DN)$ and
$\gamma N\rightarrow\bar{D}^*D^*N$, shown respectively, in Figs.
\ref{figure2} and \ref{figure3}. The results are shown in Fig. \ref{cross1}
by the solid, dotted, and dashed curves, respectively, for the reactions
$\gamma N\rightarrow \bar{D}DN$,
$\gamma N\rightarrow\bar{D}D^*N(\bar{D}^*DN)$ and
$\gamma N\rightarrow\bar{D}^*D^*N$. It is seen that the reaction
$\gamma N\rightarrow\bar{D}D^*N(\bar{D}^*DN)$ has the
largest cross section with a peak value of about 40 nb, while the reaction
$\gamma N\rightarrow \bar{D}DN$ has the smallest cross
section of only about 1 nb. The larger cross sections for processes
involving a charmed vector meson in the final state is due to the
presence of interaction vertices with three vector mesons, which
have a stronger momentum dependence than vertices with fewer number
of vector mesons, leading thus to a larger strength at high energies.

The above results are obtained without contributions from diagrams
involving the photon coupled directly to external nucleons. These
diagrams are needed to preserve the gauge invariance in each process.
Their contributions are small compared to those from diagrams with
the photon coupled directly to mesons. This is due to the $s$-channel
nucleon propagator ($1/(s-m_N^2)$) in these diagrams, which suppresses
their amplitudes more than the $t$-channel heavy meson propagator in
other as a result of the large photon energy needed to produce both the
charmed and anticharmed meson pair. In the following, we demonstrate
this effect by comparing the contribution due to diagram (1d)
with that due to diagrams (1a)-(1c) in Fig. \ref{figure1}.

The amplitude for diagram (1d) in Fig. \ref{figure1} can be written as
\begin{eqnarray}
{\cal M}&=&i2eg_{\pi NN}g_{\pi DD^*}\frac{1}{(s-m^2_N)(t-m^2_\pi)}\nonumber\\
&\times&\bar{p}(p_3)\gamma_5({p\mkern-10mu/}_1+{p\mkern-10mu/}_2+m_N)
\gamma^\mu p(p_1)\varepsilon_\mu p^\nu_5\varepsilon_\nu\nonumber\\
&\equiv&2g_{\pi DD^*}M_2p^\nu_5\varepsilon_\nu,
\end{eqnarray}
where $p_1$, $p_3$, $p_2$, and $p_5$ are the momenta of initial and final
nucleons, photon, and charmed meson, respectively. The cross section due
to this diagram alone is given by
\begin{eqnarray}
\frac{d\sigma}{dtds_1}=\frac{\sqrt{s_1}}{256\pi^2 sp^2_c}|{\cal
M}_2|^2\Gamma(s_1), \label{s-channel}
\end{eqnarray}
where again $p_c$ is the center-of-mass momentum of the nucleon
and the photon, $s_1$ is the invariant mass of the $D^{*-}D^+$
pair, and $\Gamma(s_1)$ is the decay width of the off-shell
$\pi^0\to D^{*-}D^+$.

\begin{figure}[ht]
\centerline{\epsfig{file=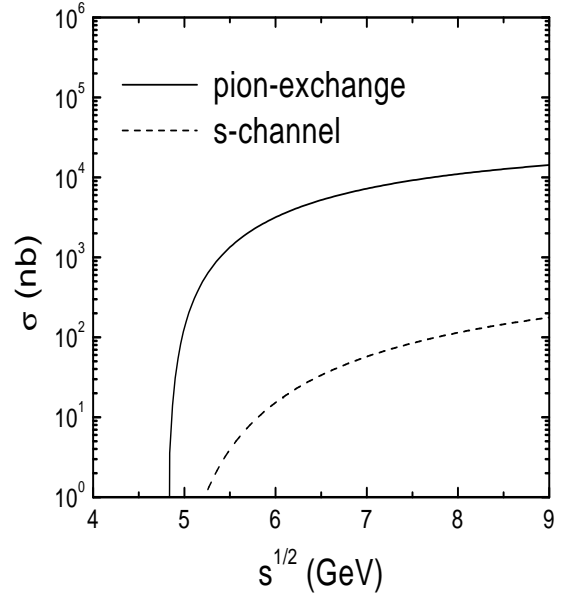,width=3.5in,height=3.5in,angle=-90}}
\caption{Cross sections due to photon coupling directly
to nucleon (diagram (1d) in Fig. \ref{figure1}, dashed curve) and due
to pion exchange (diagrams (1a)-(1c) in Fig. \ref{figure1}, solid curve).
No form factors are included at interaction vertices.}
\label{test}
\end{figure}

The cross section due to the $s-$channel diagram (1d) in Fig. \ref{figure1}
involving photon coupling directly to nucleon is shown by the dashed curve
in Fig. \ref{test} together with that coming from the pion exchange
contributions (diagrams (1a)-(1c) in Fig. \ref{figure1}), shown by the
solid curve. Form factors have been neglected in these results as we are only
interested in their relative magnitude. It is seen that the contribution
from the diagram with direct photon-nucleon coupling is more than two
orders of magnitude smaller than that from the pion-exchange diagrams and
can thus be safely neglected.

\section{Photoproduction of charmed hadrons from nucleon with two-body
final states}
\label{two}

\begin{figure}[ht]
\centerline{\epsfig{file=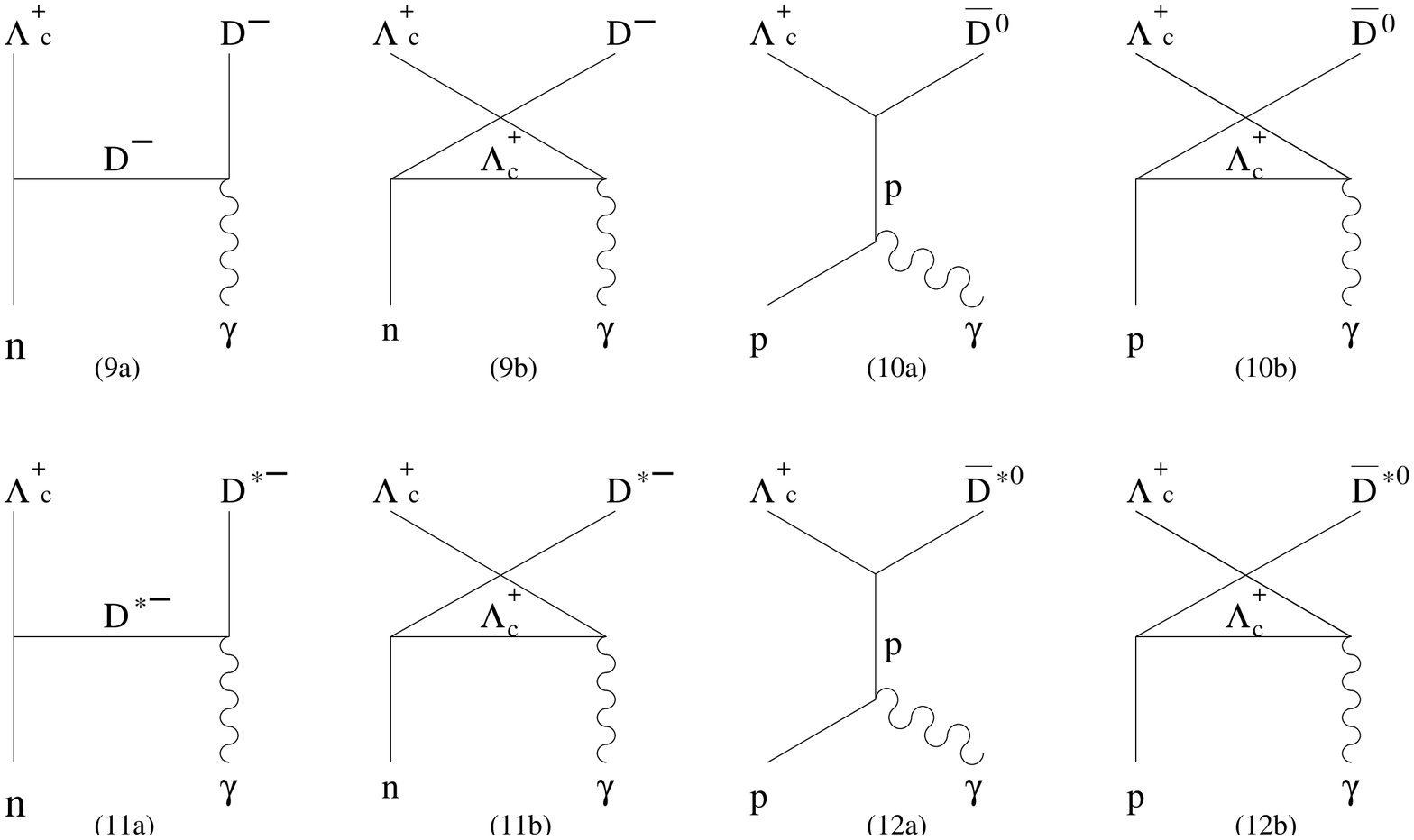,width=3in,height=2.5in,angle=0}}
\vspace{0.5cm}
\caption{Photoproduction of charmed hadrons from nucleon with
two-particle final states.}
\label{figure4}
\end{figure}

Charmed hadron can also be produced in photo-nucleon reaction from
processes involving two particles in the final state, i.e.,
$\gamma N\rightarrow\bar{D}\Lambda_c$
and $\gamma N\rightarrow\bar{D}^*\Lambda_c$, shown by diagrams in
Fig. \ref{figure4}. The interaction Lagrangians needed to evaluate
the cross sections for these reactions are:
\begin{eqnarray}
{\cal L}_{DN\Lambda_{c}} & = &ig_{DN\Lambda_{c}}
(\bar{N}\gamma_{5}\Lambda_{c}\bar{D}+D
\bar{\Lambda}_{c}\gamma_{5}N),\nonumber\\
{\cal L}_{D^{*}N\Lambda_{c}} & = &g_{D^{*}N\Lambda_{c}}
(\bar{N}\gamma_{\mu}\Lambda_{c}D^{*\mu}+\bar{D}^{*\mu}\bar{\Lambda}_{c}
\gamma_{\mu}N).
\end{eqnarray}
As in ref.\cite{liu1}, we use SU(4) relations to determine the coupling
constants $g_{DN\Lambda_c}$ and $g_{D^* N\Lambda_c}$ in terms of known
coupling constants $g_{\pi NN}$ and $g_{\rho NN}$, and they are given by
\begin{eqnarray}
g_{DN\Lambda_c}&=&\frac{3-2\alpha_D}{\sqrt{3}}
g_{\pi NN}\simeq g_{\pi NN}=13.5,\nonumber\\
g_{D^{*}N\Lambda_c}&=&-\sqrt{3}g_{\rho NN}=-5.6,
\end{eqnarray}
where $\alpha_D=D/(D+F)\simeq 0.64$ \cite{adelseck} with $D$ and
$F$ being the coefficients for the usual $D$-type and $F$-type
couplings. Values of these couplings constants are similar to those
obtained from a QCD sum rule analysis \cite{qcd}.

The amplitudes for the two processes in Fig. \ref{figure4} are
\begin{eqnarray}
{\cal M}_{9} & = & ({\cal M}^{\mu}_{9a}+{\cal M}^{\mu}_{9b})
\varepsilon_{\mu},\nonumber\\
{\cal M}_{10} & = & ({\cal M}^{\mu}_{10a}+{\cal M}^{\mu}_{10b})
\varepsilon_{\mu},\nonumber\\
{\cal M}_{11} & = & ({\cal M}^{\mu\nu}_{11a}+{\cal M}^{\mu\nu}_{11b})
\varepsilon_{\mu}\varepsilon_{\nu},\nonumber\\
{\cal M}_{12} & = & ({\cal M}^{\mu\nu}_{12a}+{\cal M}^{\mu\nu}_{12b})
\varepsilon_{\mu}\varepsilon_{\nu},
\end{eqnarray}
with ${\cal M}^{\mu}_{9a}$, ${\cal M}^{\mu}_{9b}$, ${\cal M}_{10a}$, and
${\cal M}^{\mu}_{10c}$ for the top four diagrams in Fig. \ref{figure4},
while ${\cal M}^{\mu\nu}_{11a}$, ${\cal M}^{\mu\nu}_{11b}$,
${\cal M}^{\mu\nu}_{12a}$, and ${\cal M}^{\mu\nu}_{12b}$ for
for the bottom four diagrams. They are given explicitly by
\begin{eqnarray}
{\cal M}^\mu_{9a} & = & ieg_{DN\Lambda_c}\frac{1}{t-m^2_D}(p_2-2p_4)^{\mu}
\bar{\Lambda}_c(p_3)\gamma^5 n(p_1),\nonumber\\
{\cal M}^{\mu}_{9b} & = & ieg_{DN\Lambda_c}\frac{1}{u-m^{2}_{\Lambda_c}}
\nonumber\\
&\times&\bar{\Lambda}_c(p_3)\gamma^\mu({p\mkern-10mu/}_1-{p\mkern-10mu/}_4
+m_{\Lambda_c})\gamma^5 n(p_1),\nonumber\\
{\cal M}^\mu_{10a} & = & ieg_{DN\Lambda_c}\frac{1}{s-m^2_N}\nonumber\\
&\times&\bar{\Lambda}_c(p_3)\gamma^5({p\mkern-10mu/}_1+{p\mkern-10mu/}_2+m_N)
\gamma^\mu p(p_1),\nonumber\\
{\cal M}^{\mu}_{10b} & = & ieg_{DN\Lambda_c}\frac{1}{u-m^{2}_{\Lambda_c}}
\nonumber\\
&\times&\bar{\Lambda}_c(p_3)\gamma^\mu({p\mkern-10mu/}_1-{p\mkern-10mu/}_4
+m_{\Lambda_c})\gamma^5 p(p_1),\nonumber\\
{\cal M}^{\mu\nu}_{11a} & = & eg_{D^*N\Lambda_c}\bar{\Lambda}_c(p_3)
\gamma^\alpha n(p_1)\frac{1}{t-m^2_{D^*}}\nonumber\\
&\times&\left[-g_{\alpha\beta}+\frac{(p_1-p_3)_\alpha(p_1-p_3)_\beta}
{m^2_{D^*}}\right]\nonumber\\
&\times&[2p_2^\nu g^{\beta\mu}-(p_2+p_4)^\beta g^{\mu\nu}+2p_4^\mu
g^{\beta\nu}],\nonumber\\
{\cal M}^{\mu\nu}_{11b} & = & eg_{D^*N\Lambda_c}\frac{1}{u-m^2_{\Lambda_c}}
\nonumber\\
&\times&\bar{\Lambda}_c(p_3)\gamma^{\mu}({p\mkern-10mu/}_1
-{p\mkern-10mu/}_4+m_{\Lambda_c})\gamma^\nu n(p_1),\nonumber\\
{\cal M}^{\mu\nu}_{12a} & = & eg_{D^*N\Lambda_c}\frac{1}{s-m^2_N}\nonumber\\
&\times&\bar{\Lambda}_c(p_3)\gamma^\nu({p\mkern-10mu/}_1
+{p\mkern-10mu/}_2+m_N)\gamma^\mu p(p_1),\nonumber\\
{\cal M}^{\mu\nu}_{12b} & = & eg_{D^*N\Lambda_c}\frac{1}{u-m^2_{\Lambda_c}}
\nonumber\\
&\times&\bar{\Lambda}_c(p_3)\gamma^{\mu}({p\mkern-10mu/}_1
-{p\mkern-10mu/}_4+m_{\Lambda_c})\gamma^\nu p(p_1).
\end{eqnarray}
Here, $p_1$, $p_2$, $p_3$, and $p_4$ denote the momentum of $\gamma$,
$N$, $\bar{D}(\bar{D}^*)$, and $\Lambda_c$, respectively.

The spin-averaged differential cross sections for the four
processes in Fig. \ref{figure4} are then
\begin{eqnarray}
\frac{d\sigma_{\gamma N\to\bar D\Lambda_c}}{dt} & = &\frac{1}
{256\pi sp^{2}_{c}}|{\cal M}_{i}|^{2}  |f({\bf q}^2)|^2,\label{diff1}
\label{diff4},
\end{eqnarray}
where $i=9,10,11,12$ for the the four reactions $\gamma n\to D^-\Lambda_c^+$,
$\gamma p\to \bar D^0\Lambda_c^+$, $\gamma n\to D^{*-}\Lambda_c^+$,
and $\gamma \to D^-\Lambda_c^+$.

The isospin-averaged cross sections for the
two reactions $\gamma N\to \bar{D}\Lambda_c$ and
$\gamma N\to\bar{D}\Lambda_c$ can be obtained from the above cross
sections, and they are given by
\begin{eqnarray}
\sigma_{\gamma N\to \bar{D}\Lambda_c}&=&\frac{1}{2}
(\sigma_{\gamma n\to D^-\Lambda_c^+}
+\sigma_{\gamma p\to \bar{D}^0\Lambda_c^+}),\nonumber\\
\sigma_{\gamma N\to \bar{D}^*\Lambda_c}&=&\frac{1}{2}
(\sigma_{\gamma n\to D^{*-}\Lambda_c^+}
+\sigma_{\gamma p\to\bar{D}^{*0}\Lambda_c^+}).
\end{eqnarray}

\begin{figure}[ht]
\centerline{\epsfig{file=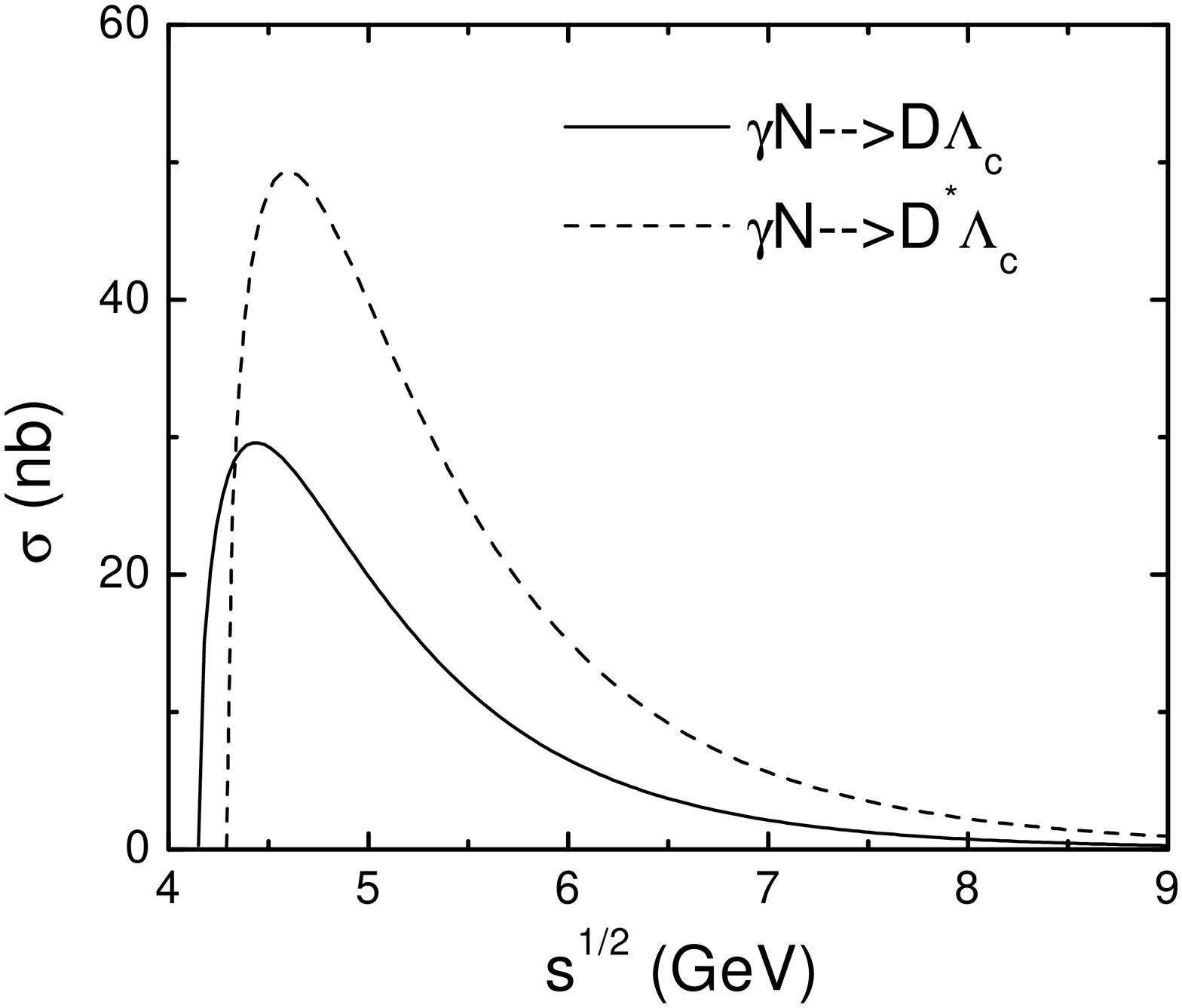,width=3.5in,height=3.2in,angle=-90}}
\caption{cross section photoproduction of charmed hadrons via charmed
meson exchange.}
\label{cross2}
\end{figure}

Taking the form factor at interaction vertices to be of the dipole type
as in Eq. (\ref{form}) and with a cutoff parameter $\Lambda=1.45$ GeV,
which is chosen to reproduce the experimentally observed relative
strength between two-body and three-body decay final states in
photoproduction of charmed hadrons on nucleon \cite{slac}, the cross
sections for the reactions $\gamma N\rightarrow\bar{D}\Lambda_c$
(dashed curve) and $\gamma N\rightarrow\bar{D}^*\Lambda_c$ (solid curve)
are shown in Fig. \ref{cross2} as functions of total center-of-mass energy.
It is seen that both cross sections are large with a peak value of about
30 nb for $\gamma N\rightarrow\bar{D}\Lambda_c$ and about 50 nb for
$\gamma N\rightarrow\bar{D^*}\Lambda_c$.

\section{Total Cross Section for charmed hadron production in
photon-nucleon reaction}\label{total}

\begin{figure}[ht]
\centerline{\epsfig{file=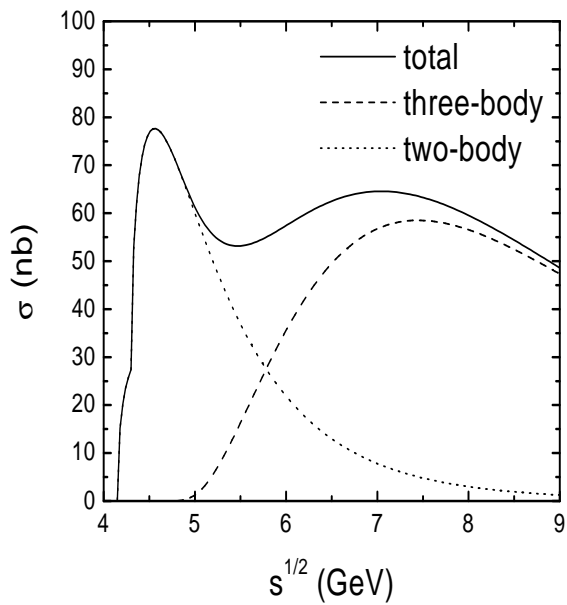,width=3.5in,height=3.5in,angle=-90}}
\caption{Total and partial cross sections for charmed hadron production in
photon-nucleon reactions as functions of center-of-mass energy.}
\label{cross3}
\end{figure}

The total cross section for photoproduction of charmed hadrons on nucleon
is given by the sum of the cross sections for two-body and three-body
final states. In  Fig. \ref{cross3}, we show the total cross section
(solid curve) together with that for two-body final state (dotted curve)
and three-body final state (dashed curve). It is seen that two-body
final states involving $\Lambda_c$ and a charmed meson dominates at
low energy, while the three-body final state involving a nucleon as well
as a charmed and anticharmed meson pair is more important at high energies.
The two have comparable magnitudes around center-of-mass energy of about
5.7 GeV. This is consistent with the experimental data at 6 GeV
\cite{slac}, which shows that the final state with a $\Lambda_c$
constitutes about 35\% of the total cross section.

Charm production from photo-nucleon reaction can be estimated
using the leading-order perturbative QCD \cite{SVZ,FS78,EN89}.
The cross section in this approach is given by
\begin{eqnarray}
\sigma^{\gamma N} (\nu)=\int_{2m_c^2/\nu}^1 dx ~
\sigma^{\gamma g}(\nu x) g(x),
\end{eqnarray}
where $m_c$ is the charm quark mass,
$g(x)$ is the gluon distribution function inside the nucleon,
and $\nu=p\cdot p_\gamma$ with $p$ and $p_\gamma$ being the momenta of
the incoming nucleon and photon.  The cross section
$\sigma^{\gamma g}(\omega)$ is that for charm-anticharm quark production
from the leading order photon-gluon scattering, i.e.,
\begin{eqnarray}
\sigma^{\gamma g \rightarrow \bar{c} c}(\omega)& =&
\frac{ 2 \pi \alpha_s \alpha}{9} \frac{4}{\omega^2}
\bigg[ \bigg( 1 +\frac{4m^2}{\omega^2}-\frac{8m_c^2}{\omega^4}\bigg)
\nonumber\\
&\times&{\rm log}\frac{1+\sqrt{1-\frac{4m_c^2}{\omega^2}}}
{1+\sqrt{1-\frac{4m_c^2}{\omega^2}}}\nonumber\\
&-&\bigg( 1 +\frac{4m_c^2}{\omega^2} \bigg)\sqrt{1-\frac{4m_c^2}{\omega^2}}
\bigg],
\end{eqnarray}
where $\omega^2=2 p_g\cdot p_\gamma$, with the gluon momentum
denoted by $p_g$.

\begin{figure}[ht]
\centerline{\epsfig{file=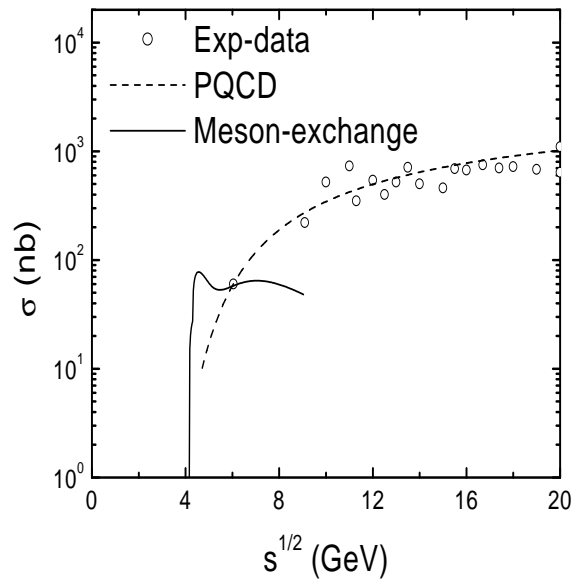,width=3.5in,height=3.5in,angle=-90}}
\caption{Cross sections for charm production in photo nucleon reaction
in the hadronic model (solid curve) and the pQCD approach
(dashed curve). The experimental data are shown by open circles.}
\label{cross4}
\end{figure}

Using $m_c=1.3$ GeV and the leading order MRST 2001 parameterization
of the gluon distribution function in nucleon \cite{mrst}, we have
calculated the photo charm production cross section on nucleon using
the LO QCD formula, and the result is shown Fig. \ref{cross4} by
the dashed curve together with that from the effective hadronic model
(solid curve) and the available experimental data (open circles).
We see that the LO QCD result reproduces the data at 6 GeV and
at higher energies.  However, the QCD prediction below 6 GeV falls
well below the results from the effective hadronic model. It is known
that the QCD formula for photoproduction of heavy quarks should work
best when the momenta involved are of order the heavy quark mass $m_c$.
Below this momentum and near the threshold energy, large logarithms will
appear in the perturbative QCD approach and spoil its convergence
\cite{Frixione}.  At these energies, our phenomenological hadronic
approach should be more reliable as the cross section is dominated
by two-body final states with no additional contribution to cause any
large correction. On the other hand, the results from the hadronic model
at higher energies fall short of the experimental data. This is expected
because at these energies, contributions from four-body final state and
from the exchange of heavier mesons will become important. At these
higher energies, perturbative QCD calculations should be a more efficient
way for determining the cross section for photo charm production than
adding new ingredients into our phenomenological hadronic model.

\section{summary and discussions}\label{conclusion}

In summary, the total cross section for charmed hadron production in
photo-nucleon reaction is evaluated in an effective hadronic model.
This model is based on a gauged SU(4) flavor symmetric Lagrangian
with the photon introduced as a $U_{\rm em}(1)$ gauged particle. The
symmetry breaking effect is taken into account via using empirical
hadron masses and coupling constants. Form factors of the monopole
type are introduced at interaction vertices. This hadronic model
has been previously used to evaluate the dissociation cross section
of $J/\psi$ by hadrons. For photo production of charmed hadrons on
nucleon, we have included both two-body final states involving a
$\Lambda_c$ and a charmed meson as well as three-body
final states involving a nucleon and a charmed and anticharmed meson
pair. It is found that reactions with two-body final states dominate
the production cross section at low center-of-mass energies while reactions
with three-body states are more important at high center-of-mass energies.
Using cut-off parameters in the form factors from previous studies of
$J/\psi$ absorption by hadrons, the model reproduces the lowest available
experimental data at center-of-mass energy of 6 GeV. Our results thus
again confirm the validity of the effective hadron model in previous
studies and provide a useful model for further studies of reactions
involving heavy quarks at low and near threshold energies.

\section*{acknowledgment}

This paper is based on work supported by the National Science Foundation
under Grant No. PHY-0098805 and the Welch Foundation under Grant No. A-1358.
SHL is also supported in part by the KOSEF under Grant No. 1999-2-111-005-5
and by the Korea Research Foundation under Grant No. KRF-2002-015-CP0074.

\end{multicols}

\end{document}